\documentstyle[12pt,epsfig]{article}   

\def\R{ {\rm R \kern -.31cm I \kern .15cm}}
\def\C{ {\rm C \kern -.15cm \vrule width.5pt \kern .12cm}} 
\def\Z{ {\rm Z \kern -.27cm \angle \kern .02cm}} 
\def\N{ {\rm N \kern -.26cm \vrule width.4pt \kern .10cm}} 
\def\1{{\rm 1\mskip-4.5mu l} }
\def\lsim{\raise0.3ex\hbox{$<$\kern-0.75em\raise-1.1ex\hbox{$\sim$}}}
\def\gsim{\raise0.3ex\hbox{$>$\kern-0.75em\raise-1.1ex\hbox{$\sim$}}}
\def\noi{\noindent} 

\def\beq{\begin{equation}}   
\def\eeq{\end{equation}}
\def\bea{\begin{eqnarray}}  
\def\eea{\end{eqnarray}} 
\def\nn{\nonumber}
\def\noi{\noindent} 
\def\beeq{\begin{eqnarray}}
\def\eeeq{\end{eqnarray}}
\newcommand\mysection{\setcounter{equation}{0}\section}
\renewcommand{\theequation}{\thesection.\arabic{equation}}
\newcounter{hran} 
\renewcommand{\thehran}{\thesection.\arabic{hran}}

\def\bmini{\setcounter{hran}{\value{equation}}
  \refstepcounter{hran}\setcounter{equation}{0}
  \renewcommand{\theequation}{\thehran\alph{equation}}\begin{eqnarray}}

\def\bminiG#1{\setcounter{hran}{\value{equation}}
\refstepcounter{hran}\setcounter{equation}{-1}
\renewcommand{\theequation}{\thehran\alph{equation}}
\refstepcounter{equation}\label{#1}\begin{eqnarray}}

\def\emini{\end{eqnarray}\relax\setcounter{equation}{\value{hran}}\renewcommand{\theequation}{\thesection.\arabic{equation}}}

\begin{document} 
\vbox to 2 truecm {}  
\centerline{\Large\bf Brane Universes, AdS/CFT, Hamiltonian Formalism}  
\vskip 3 truemm
\centerline{\Large\bf  and the Renormalization Group\footnote{Based on
a lecture at the LPT Orsay}} \vskip 1 truecm

\vskip 1 truecm
\centerline{\bf Ulrich Ellwanger} 
\centerline{Laboratoire de Physique
Th\'eorique\footnote{Unit\'e Mixte de Recherche - CNRS - UMR 8627}} 
\centerline{Universit\'e de Paris XI, B\^atiment 210, F-91405 ORSAY
Cedex, France} \vskip 2 truecm

\begin{abstract} The AdS/CFT correspondence is developed from classical
solutions on AdS$_5$ with two boundaries. The corresponding limits and
the reduction of degrees of freedom are discussed, as well as  the
required renormalization on the field theory side. The Hamiltonian
first-order approach towards the solution of coupled
gravitational/matter equations of motion is introduced, and the RG
interpretation is exposed. Finally we discuss a recent approach towards
a naturally vanishing cosmological constant which is based on the
AdS/RG correspondence. \end{abstract} 

\vskip 3 truecm \noi LPT Orsay 00-64 \par \noi July 2000 \par

\newpage \pagestyle{plain} \baselineskip 18pt

\mysection{Introduction} 
\hspace*{\parindent} Recently two conceptually
different applications of 5-dimensional universes supplemented with (3
+ 1)-dimensional branes have been intensively discussed:\par

 i) Brane universes, where the 5th dimension is a physical (spatial)
dimension; the study of its physical consequences resembles to the
standard Kaluza-Klein reduction (from 5 to 4 dimensions): One
decomposes the fields into modes which solve the equations of motion
(including contributions from the actions on the branes) along the 5th
dimension. Of particular interest are the gravitational modes and the
induced cosmological evolution. \par

ii) the AdS/CFT-correspondence, which relates - via a holographic
principle - classical solutions of a (super-) gravity theory on AdS$_5$
to correlation functions of composite operators of a (3 + 1)
dimensional quantum field theory. \par

The aim of the present note is a discussion of the conceptual
differences and common techniques of both approaches. It is based on a
lecture at the LPT Orsay. \par

First, in section 2, we consider the general solution of the equations
of motion of a (free massive) scalar field in a 5-dimensional bulk with
(negative) cosmological constant, i.e. AdS$_5$. (The corresponding
Einstein-equations are not re-derived; to this end we refer to, e.g.,
\cite{BD1,RS1,RS2}. The boundary or ``jump'' conditions at branes do,
however, not yet play a role). Following, to a large extent, \cite{GK}
we consider the classical action integrated over a part of AdS$_5$. The
aim is to see the reduction of ``degrees of freedom'', i.e. integration
constants, in the limit where the integral extends far inside of
AdS$_5$. One recovers the holographic principle, a one-to-one
correspondence between the integrated classical action and correlators
of a (3+1)-dimensional QFT \cite{M1,W1,GK}. The required limits,
associated to renormalization on the QFT side, are discussed in some
detail. We add a very superficial mini-review on recent applications of
the AdS/CFT-correspondence - incomplete and without references.  \par

The next chapter (section 3) is dedicated to the Hamiltonian formalism.
Evidently both 5-d cosmology and AdS/CFT-correspondence require
solutions of the (classical) combined gravitational and matter
equations of motion. We discuss the Hamiltonian first order approach,
an identity for the $y$-integrated bulk action, and the Hamiltonian
constraint, valid in the presence of gravity. This section is
essentially based on \cite{B1} (see also \cite{F1}). Following
\cite{B1,V2,V3,V1} we draw the analogy to renormalization group (RG)
equations. \par

In the last section 4 we comment on some recent proposals towards a
vanishing cosmological constant without fine tuning. We discuss, in
particular, the approach of \cite{V2,V1,V3} based on the RG
interpretation of the 5-d dynamics - at least the way we understand it.
\par

The purpose of this paper is purely pedagogical. Its aim is to discuss
the interface between 5-d cosmology and the AdS/CFT-correspondence; it
is not meant to represent a review on any of these two subjects. Hence
the corresponding discussions and references are far from complete. We
found it appropriate, on the other hand, to discuss the links between
these subjects employing common conventions and common techniques.

\mysection{CFT from ADS$_5$ with two boundaries} 
\hspace*{\parindent}
Let us start, to fix the conventions, with pure gravity in a
5-dimensional bulk with a negative cosmological constant. The
corresponding action reads (with $R_{\mu \nu} = R^\rho_{\ \mu \nu \rho}$)  

\beq  \label{2.1e} S_{Grav} = - \int dy \int d^4x \sqrt{-g_5} \left \{
{1 \over 2 \kappa_5^2} R - \Lambda \right \} + \hbox{Boundary Terms}
\quad .\eeq

The curvature scalar $R$ contains second derivations of the metric;
before deriving the gravitational equations of motion, these terms
should be transformed into expressions quadratic in first derivatives
by means of partial integrations. The boundary terms in (\ref{2.1e})
are chosen such that they cancel precisely the corresponding total
derivative terms \cite{G1}. \par

We look for a 5-dimensional metric, which solves the Einstein
equations following from (\ref{2.1e}) and preserves 4-dimensional
Poincar\'e-invariance, of the form

\bminiG{2.2e} \label{2.2ae} ds^2 = a^2(y) \eta_{ij} dx^i dx^j +
b^2(y) \ d^2y \quad , \eeeq  \beeq \eta_{ij} dx^i 
dx^j = d \vec{x}^{\, 2} - dt^2 \quad . \label{2.2be} \emini

\noi $i$, $j$ are indices perpendicular to the 5th dimension and take
the four values $0 \dots 3$; greek indices $\mu$, $\nu$ will take all
five values $0 \dots 3$ and 5. \par

The following expressions for $a(y)$ and $b(y)$, which solve the
Einstein equations and describe an AdS space, are most frequently
used:

\bminiG{2.3e} 
\label{2.3ae} a(y) = e^{-y / \lambda} \quad , \quad
b(y) = 1 \qquad \hbox{with} \quad \lambda^2 = {6 \over \Lambda
\kappa_5^2} \quad ,  \eeeq \noi or, with $y' = \lambda \exp
(y/\lambda)$ and omitting the prime,  \beeq a(y) = b(y) = {\lambda
\over y} \quad . \label{2.3be} \emini

\noi Often the convention $\lambda = 1$ is employed. \par

In the case of (\ref{2.3ae}) $a(y)$ diverges for $y \to - \infty$,
whereas for (\ref{2.3be})  $y$ takes only positive values and $a(y)$
diverges at $y = 0$. In both cases the ``warp factor'' $a(y)$ decreases
for increasing $y$. (Sometimes, however, $y$ is replaced by $- y$ in
(\ref{2.3ae})).
\par

In the Randall-Sundrum model I \cite{RS1} two flat 3-branes are placed
into the bulk: Using the metric (\ref{2.3ae}), brane 1 (with positive
tension) is situated at $y_1 = 0$, and brane 2 (with negative tension)
at $y_2 = \pi r_c$. ``Standard matter fields'' are supposed to live on
brane 2 where the warp factor is exponentially small compared to brane
1. The orbifold geometry $S^1/Z_2$ (with $a(y)$, $b(y)$ even) can be
represented on the entire $y$ axis with $a(-y) = a(y)$ , $a(y + 2n \pi
r_c) = a(y)$ ($n$ integer), and similarly for $b(y)$ and all other
(even) fields. \par

In the Randall-Sundrum model II \cite{RS2} ``standard matter fields''
are supposed to live on brane 1, whereas the location of the brane 2 is
pushed to $y_2 \to + \infty$. \par

The first scenario is depicted in Fig. 1, where we plot the warp
factor $a(y)$ versus $y$ for the metric (\ref{2.3ae}). The branes are
represented by vertical lines in the positive or negative direction
according to their tension. The behavior of $a(y)$ for $y < y_1$ or $y
> y_2$, in the case of an orbifold geometry, is indicated as dashed
lines. \par

\begin{figure}[p] \unitlength1cm \begin{picture}(17,17)
\put(-3.0,-8.5){\epsfig{file=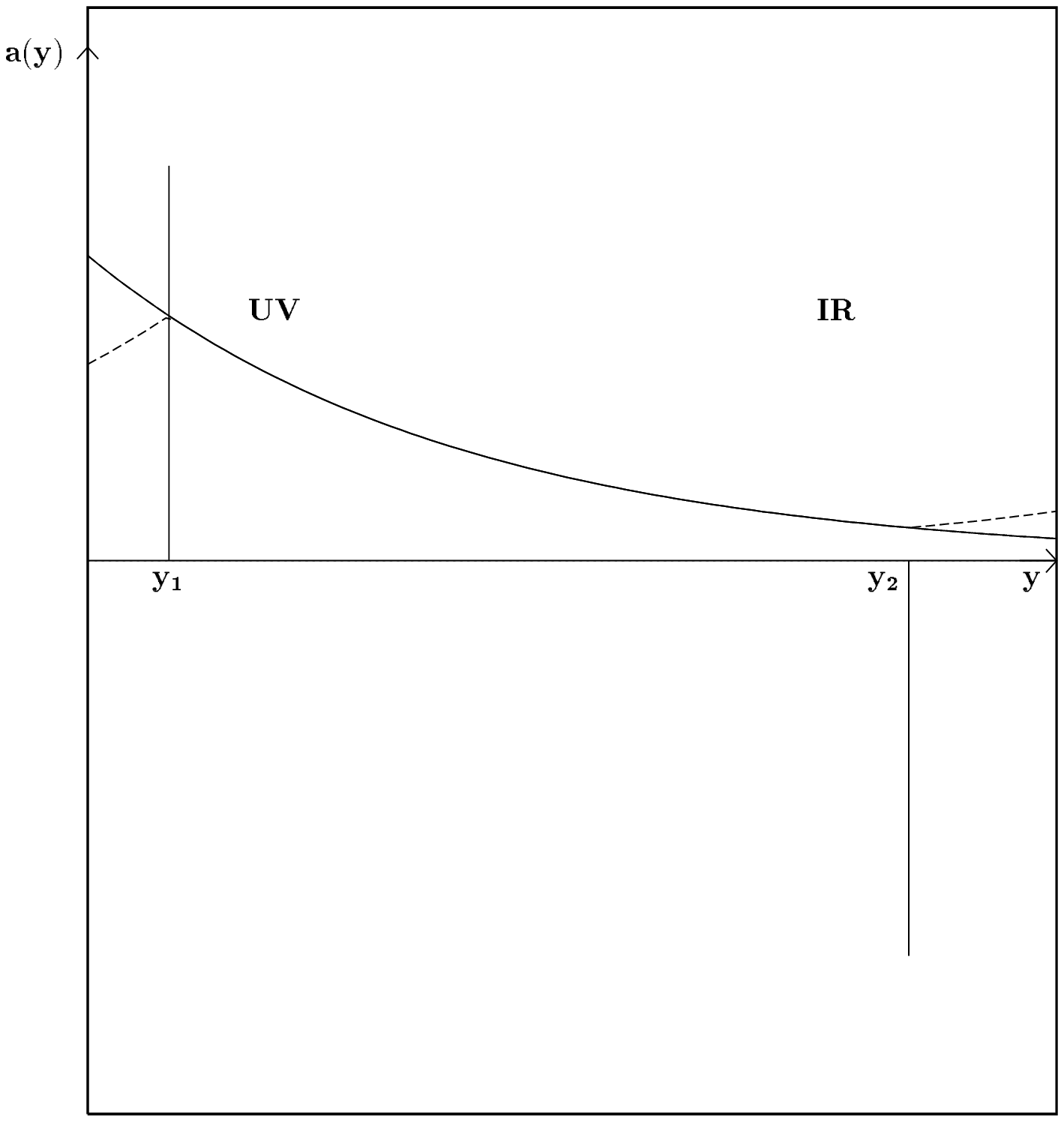}} 
\end{picture} {\bf Fig. 1:} The warp factor $a(y)$ versus $y$ for the
metric (\ref{2.3ae}). The branes are represented by vertical lines in
the positive or negative direction according to their tension. The
behavior of $a(y)$ for $y < y_1$ or $y > y_2$, in the case of an
orbifold geometry, is indicated as dashed lines. "UV" and "IR" are the
corresponding regimes of a 4-d field theory.

\end{figure}

Now we will add matter to the bulk, first in the form of a free
massive scalar field $\varphi$. The action reads

\bea \label{2.4e} &&S = S_{Grav} + S_{matter} \quad , \nn \\
&&S_{matter} = - \int dy \int d^4x \sqrt{-g_5} \left \{ {1 \over 2}
\partial_{\mu} \varphi \ g^{\mu \nu} \ \partial_{\nu} \varphi + {1
\over 2} m^2 \varphi^2 \right \} \quad , \eea

\noi and the scalar equation of motion is given by

\beq \label{2.5e} \left ( \partial_{\mu} \sqrt{-g_5} \ g^{\mu \nu}
\partial_{\nu} - \sqrt{-g_5} \ m^2 \right ) \varphi = 0 \quad .  \eeq

\noi Subsequently it is somewhat more convenient to employ the metric
(\ref{2.3be}) where eq. (\ref{2.5e}) turns into

\beq \label{2.6e} \left ( \partial_y^2 - {3 \over y} \partial_y +
\eta^{ij} \partial_i \partial_j - {\lambda^2 m^2 \over y^2} \right
) \varphi (y, x^i ) = 0 \quad . \eeq

\noi Next we consider the Fourier transform in the 4-dimensional space
$\{x^i\}$ per\-pen\-di\-cu\-lar to $y$: 

\beq \label{2.7e} \varphi (y, x^i) = \int {d^4p \over (2 \pi )^4} \
e^{ip\cdot x} \ \phi (y, p) \eeq

\noi whereupon the equation of motion becomes

\beq \label{2.8e} \left ( \partial_y^2 - {3 \over y} \partial_y - \left
( p^2 + {\lambda^2 m^2 \over y^2}\right ) \right ) \ \phi (y,p) = 0
\eeq

\noi with $p^2 = \vec{p}^{\, 2} - p_0^2$. \par

As a second order differential equation in $y$ the general solution
admits two ($p$-dependent) integration constants $C_1$, $C_2$, and
reads

\beq \label{2.9e} \phi (y, p) = C_1(p) \ y^2 \ K_{\nu}(py) + C_2 (p) \
y^2 \ I_{\nu}(py) \eeq

\noi with $\nu = \sqrt{4 + \lambda^2 m^2}$, $p = \sqrt{p^2}$, and where
$K_{\nu}$ and $I_{\nu}$ are Bessel functions. As boundary conditions we
can thus impose the values of $\phi$ at two different values of $y$ as

\beq \label{2.10e} \phi (y_1, p) \mathrel{\mathop =^{!}} \phi_1 (p)
\quad , \qquad \phi(y_2,p) \mathrel{\mathop =^{!}} \phi_2(p) \quad
.\eeq

\noi These boundary conditions fix $C_1(p)$, $C_2(p)$ to be of the form

\bea 
\label{2.11e} 
&&C_1(p) = {1 \over N} \left ( y_1^{-2} \ \phi_1(p) \ I_{\nu}(py_2) -
y_2^{-2} \ \phi_2 (p)\  I_{\nu}(py_1) \right ) \quad , \nn \\ 
&&C_2(p) = {1 \over N} \left ( y_2^{-2} \ \phi_2 (p) \ K_{\nu}(py_1) -
y_1^{-2} \ \phi_1(p) \ K_{\nu} (py_2) \right ) \quad , \nn \\ 
&&N = K_{\nu}(py_1) \ I_{\nu}(py_2) - K_{\nu}(py_2) \
I_{\nu}(py_1) \quad .\eea

\noi The next object of our desire is the matter action (\ref{2.4e}),
with the $y$ integral confined to the interval $\{y_1, y_2\}$:

\beq \label{2.12e} S_{1,2} = - \int_{y_1}^{y_2} dy \int d^4x
\sqrt{-g_5} \left \{ {1 \over 2} \partial_{\mu} \varphi \ g^{\mu \nu} \
\partial_{\nu} \varphi + {1 \over 2} m^2 \ \varphi^2 \right \} \eeq 

\noi where $\varphi$ is a solution of the equations of motion. Using
again the metric (\ref{2.3be}), the Fourier transform (\ref{2.7e}) and
the equations of motion (\ref{2.8e}), $S_{1,2}$ can be expressed as 

\beq \label{2.13e} S_{1,2} = - \left [ {\lambda^3 \over 2y^3} \int
{d^4p \over (2 \pi )^4} \ \phi (y, p) \ \partial_y \phi(y,-p) \right
]_{y_1}^{y_2} \quad . \eeq

\noi Clearly, with $\phi$ given by (\ref{2.9e}) and the integration
constants given by (\ref{2.11e}), $S_{1,2}$ will be a (quadratic)
functional of the boundary values $\phi_1(p)$ and $\phi_2(p)$. It can
be straightforwardly obtained from (\ref{2.13e}), but its explicit
expression is somewhat lengthy. \par

Now we want to push $y_2 \to + \infty$, keeping $\phi_2(p)$ finite.
 From $I_{\nu}(py) \sim (2 \pi py)^{-{1 \over 2}}
e^{py}$ for $y \to \infty$ one finds, from (\ref{2.11e}), that $C_2(p)$
vanishes. Alternatively, from (\ref{2.9e}) one deduces that $C_2(p)$ has
to vanish, if $\phi (y, p)$ is required to be bounded for $y \to +
\infty$. Hence the expression (\ref{2.9e}) for $\phi (y, p)$
assumes the form

\beq \label{2.14e} \phi (y, p) = {y^2K_{\nu}(py) \over y_1^2 K_{\nu}
(py_1)} \ \phi_1(p) \quad , \eeq 

\noi i.e. all dependence on $\phi_2(p)$ has disappeared. \par

This is a version of the celebrated holographic principle: The space
under consideration is AdS$_5$ with an inner boundary located at $y =
y_1$ (the part of the space with $y < y_1$ is cut off), and the
configuration of $\phi (y, p)$, for all $y \geq y_1$, is determined by
its value on $\phi_1(p)$ on the boundary. We have ``lost'' one
integration constant of the equations of motion through the requirement
that $\phi (y, p)$ is bounded all over AdS$_5$ within the boundary,
notably for $y \to + \infty$. \par

Denoting $S_{1,2}$, for $y_2 \to \infty$, by $S_1$, we obtain from
(\ref{2.13e}) and (\ref{2.14e}) (using a recursion formula for
$K_{\nu}$) 

\beq \label{2.15e} S_1(\phi_1) = - {\lambda^3 \over 2 y_1^4} \int {d^4p
\over (2 \pi )^4} \ \phi_1(p) \ \phi_1(-p) \ \left ( - 2 - \nu + {py_1
\ K_{\nu+1}(py_1) \over K_{\nu} (py_1)} \right ) \quad . \eeq

The next steps are the following: we are interested in the limit $y_1
\to 0$ and, most importantly, we interpret $S_1(\phi_1)$ 
differently: we identify $\phi_1$ (or its Fourier transform $\varphi_1
(x^i)$) with a source for a (composite) operator ${\cal O}$ of some
4-dimensional field theory, and $S_1(\phi_1)$ (or $S_1(\varphi_1)$) as
the corresponding generating functional of connected Green functions
\cite{W1,GK}. In the Euclidean 4-d theory we assume

\beq \label{2.16e} e^{-S_1(\varphi_1)} = <e^{\varphi_1 \cdot {\cal
O}}> \equiv \int {\cal D} \chi \ e^{-S(\chi )+ \int \varphi_1 {\cal O}
(\chi ) d^4 x} \eeq

\noi where $S(\chi )$ is the action of some conformal field theory with
fundamental fields $\chi$, and ${\cal O} (\chi )$ a composite operator.
Conventionally sources for operators are denoted by $J$, and one writes

\beq \label{2.17e} e^{-G(J)} = \int {\cal D} \chi \ e^{-S(\chi )
+ \int J {\cal O} (\chi ) d^4x} \quad . \eeq

\noi The statement thus reads $S_1(\varphi_1 ) = G(J = \varphi_1 )$.
\par

A remark on the renormalization of the generating functional of Green
functions of composite operators is in order: Now, in general,
multiplicative renormalization of $J$ (or the operator ${\cal O} (\chi
)$) is not sufficient to render Green functions with several
insertions of ${\cal O} (\chi )$ -~i.e. higher powers of $J$ in
$G(J)$~- UV finite. To this end one has to add a local polynomial in
$J$ and derivatives (with divergent coefficients in the limit where a
UV cutoff is removed) to $G(J)$ (see, e.g., \cite{Col}).
Perturbatively, this polynomial is of finite order in $J$ and
derivatives, if the operator ${\cal O} (\chi )$ is relevant (if the
mass dimension of $J$ is positive), but of infinite order in $J$
otherwise. \par

Let us return to the $y_1 \to 0$ limit of $S_1(\phi_1)$ in
(\ref{2.15e}), which is obviously divergent. Within the interpretation
(\ref{2.16e}) of $S_1(\phi_1)$ it is natural to relate
$y_1^{-1}$ to some UV cutoff $\Lambda_{UV}$ of the 4-dimensional field
theory. In view of the invariance of the 5-dimensional classical theory
(including gravity) under coordinate reparametrizations it is wiser,
however, to associate $\Lambda_{UV}$ directly to the warp factor
$a(y)$: 

\beq \label{2.18e} \Lambda_{UV} \sim a(y_1)\cdot M \eeq    

\noi where $M$ is some fundamental scale. (Within the present metric
(\ref{2.3be}) we have, of course, $a(y_1) \sim y_1^{-1}$). \par

In order to ``renormalize'' $S_1(\phi_1)$ to the order $\phi_1^2$ we
are thus allowed to add polynomials in $p^2$ with divergent
coefficients for $y_1 \to 0$. Let us assume that $\nu$ equals an
integer $n$, at it happens in most cases (see below). The leading
behavior of $K_n(z)$ for $z \to 0$ reads

\bea \label{2.19e} K_n(z) &=& \sum_{k=0}^{n-1} \ {(-1)^k \over k !} \
2^{2k-n-1} \ \Gamma (n - k) \ z^{2k - n} \nn \\ &&+ (-1)^{n+1} \
\sum_{k=0}^{\infty} \ {2^{-2k-n} \over k! \Gamma (n+k+1)} \ z^{2k+n} \
(\ln z + \hbox{const.}) \quad .  \eea

\noi Due to the presence of the logarithm in (\ref{2.19e}) we cannot
cancel all terms in $S_1(\phi_1)$, for $y_1 \to 0$, by a polynomial in
$p^2$. If we rewrite $S_1(\phi_1)$ in the form 

\beq \label{2.20e} S_1(\phi_1) = - {1 \over 2} \int {d^4p \over ( 2 \pi
)^4} \ \phi_1(p) \ \phi_1(-p) \ F(y_1, p^2) \eeq  

\noi the leading non-analytic term in $F(y_1,p^2)$ reads \cite{GK}

\beq \label{2.21e} F(y_1, p^2) \sim \lambda^3 y_1^{2n-4} \ {2^{1-2n}
\over ((n-1)!)^2} \ (-p^2)^n \ \ln p^2 \quad . \eeq

\noi For $n \not= 2$ the remaining dependence on $y_1$ can be cancelled
by multiplicative renormalization: 

\beq \label{2.22e} \phi_1 = y_1^{2-n} \ \phi_{1,ren} \eeq  

\noi Within the interpretation (\ref{2.16e}), and after
Fourier-transformation back to ordinary space, (\ref{2.20e}) with
(\ref{2.21e}) implies the following behaviour of the 2-point-function of
the operator ${\cal O}$ associated to the ``source'' $\phi_1$: 

\beq \label{2.23e} <{\cal O} (0) \ {\cal O} (x)> \sim {1 \over
|x|^{2\Delta}} \quad , \quad \Delta = 2 + n \quad . \eeq

\noi This is indeed the behaviour of a 2-point-function of an operator
of dimension $\Delta$ in a conformal field theory. (This result can
also be derived directly in ordinary space, for arbitrary $\nu \not=$
integer, provided the required ``renormalization'' is carefully
performed \cite{W1}). The fact that (\ref{2.23e}) (and, by the way, the
correct power of $y_1$ in (\ref{2.22e}), if expressed in terms of
$\Delta$) relate properties of a 4-dimensional (conformal) quantum
field theory to classical solutions of a 5-dimensional field theory on
AdS is the ``AdS/CFT-miracle''. \par

Albeit the purpose of the present note is to discuss relations between
4d/5d field theories, we will make some brief remarks on its
10-dimensional origin. The underlying framework considered by Maldacena
\cite{M1} was type II B string theory with $N$ D3-branes on top of each
other; this gives rise to a 10-dimensional metric of the form

\beq \label{2.24e} ds^2 = \left ( 1 + {y^4 \over \lambda^4} \right
)^{-{1 \over 2}} \ \eta_{ij} \ dx^i \ dx^j + \left ( 1 + {y^4 \over
\lambda^4} \right )^{1 \over 2} \left ( {\lambda^4 \over y^4} \ dy^2 +
{\lambda^4 \over y^2} \ d\Omega_5^2 \right ) \quad . \eeq

\noindent $d\Omega_5^2$ is the metric of $S_5$. $\lambda$ is related to
the gauge coupling $g_{YM}$ of a $SU(N)$ ${\cal N} = 4$ super $YM$
theory (which appears as the 4-dimensional conformal field theory in
this case) and the string scale $\alpha '$ via

\beq \label{2.25e} \lambda^4 = 2N g_{YM}^2 \ \alpha '^{\, 2} \eeq

\noi in the t'Hooft limit $Ng_{YM}^2$ large (but $\alpha '$ is small).
In the limit

\beq \label{2.26e} {y^2 \over \lambda^2} \gg 1 \eeq

\noi the metric (\ref{2.24e}) simplifies to 

\beq \label{2.27e} ds^2 = {\lambda^2 \over y^2} \left ( \eta_{ij} \
dx^i \ dx^j + dy^2 \right ) + \lambda^2 d\Omega_5^2 \quad . \eeq 

\noi First, the part depending on $x^i$, $y$ coincides with the AdS$_5$
metric (\ref{2.3be}). Second, the radius $\lambda$ of $S_5$ has become
$y$-independent. The 5-dimensional theory on AdS$_5$ is gauged ${\cal
N} = 8$ supergravity \cite{PG}, supplemented with Kaluza-Klein modes
from $S_5$. Since the masses of the Kaluza-Klein modes are
corresponding multiples of the inverse radius $\lambda$ (= $\lambda$ as
defined in (\ref{2.3ae})) one finds indeed that $\nu$, as defined below
(\ref{2.9e}), assumes integer values in these cases (as used in
(\ref{2.19e}), (\ref{2.21e})). \par

Since, for $\lambda$ fixed, we have

\beq \label{2.28e} \alpha ' = {\lambda^2 \over \sqrt{2N g_{YM}^2}} \eeq

\noi one finds that higher orders in the inverse t'Hooft coupling
$Ng_{YM}^2$ correspond to higher orders in the string scale $\alpha '$.
\par

In view of the limit (\ref{2.26e}) one may wonder why the AdS/CFT
correspondence, which seemed to be based on the limit $y \to 0$, works
at all. However, the crucial point is not the limit $y \to 0$, but the
``reduction of the degrees of freedom'', i.e. the reduction of
integration constants of an arbitrary bulk solution (\ref{2.9e}). The
constraint $C_2(p) = 0$ in (\ref{2.9e}) arose from pushing the brane on
$y_2$ to $y_2 \to + \infty$, and from requiring the bulk solution $\phi
(y, p)$ to remain bounded for $y \to + \infty$. It is thus the large
$y$ regime (satisfying (\ref{2.26e})) which leads to the holographic
principle, the one-to-one correspondence of a classical bulk solution
to a configuration on a brane at $y_1$. The replacement of the Bessel
function $K_n(z)$ by its leading behaviour for $z \to 0$ in
(\ref{2.19e}) requires $y^2p^2 \ll 1$ instead of (\ref{2.26e}); taken
together both inequalities require

\beq \label{2.29e} p^2 \ll {1 \over \lambda^2} \quad .  \eeq

\noi For momenta violating (\ref{2.29e}) higher derivative terms in the
supergravity action (associated to higher orders in the string scale
$\alpha '$) would become relevant. \par

Let us very briefly -~without references~- and without pretension to
completeness list scenarios to which the 5d/4d correspondence has been
applied. First, one can use the trivial vacuum of ${\cal N} = 8$
supergravity on AdS$_5$ (with a $y$-independent cosmological constant)
as a background. On this background one can consider $x^i$-dependent
fluctuations of the various fields, i.e. solve the equations of motion
(with boundary conditions at $y_1$, which fix the bulk solutions
uniquely) iteratively to 2nd, 3rd or even 4th order in the fields. The
corresponding integrated action $S_1(\phi_1)$ then allows to obtain 2,
3 or even 4 point functions of composite operators in $d = 4$ ${\cal N}
= 4$ $SU(N)$ Yang-Mills theory. \par

The various scalars of ${\cal N} = 8$ supergravity -~including
Kaluza-Klein modes~- can indeed be interpreted as sources for (gauge
invariant) local operators of ${\cal N} = 4$ $SU(N)$ $YM$. The various
vector fields $A_{\mu}$ of ${\cal N} = 8$ supergravity (for $\mu \not=
5$) can be interpreted as sources for currents $J_{\mu}$, associated to
global symmetries of ${\cal N} = 4$ $SU(N)$ $YM$. Fluctuations of the
traceless components of the graviton (with $\mu$, $\nu \not= 5$) give
correlators of the energy-momentum tensor of ${\cal N} = 4$ $SU(N)$
$YM$; the two-point function allows to obtain the analog of a central
charge and a $c$-theorem. Fluctuations of the dilaton give correlators
of the trace of the energy-momentum tensor. \par

Second, one can look for a non-trivial background solution of $d = 5$
${\cal N} = 8$ supergravity: A solution of the scalar equations of
motion and the Einstein equations with a $y$-dependent
($x^i$-independent) metric and (some) $y$-dependent ($x^i$-independent)
scalar fields. The AdS$_5$/CFT-correspondence is maintained, if i) for
small $y$ (large $a(y)$) the scalar fields are in the trivial vacuum;
ii) for large $y$ the scalars (and hence $a(y)$) assume a constant,
$y$-independent value. Hence 5-dimensional space-time is of the
AdS$_5$-form both for small and large $y$, but, in general, with
different cosmological constants. In order to solve the scalar
equations of motion, the values of the scalar fields are at extrema of
the scalar potential both at small $y$ (trivially) and at large $y$.
These configurations are called kink-solutions. One can show that
generally $a(y)$ still decreases in $y$ for all $y$. \par

 From the point of view of a 4-dimensional field theory at fixed $y$,
the scalar fields are sources for composite operators or, since the
sources are $x^i$-independent, masses and couplings. The $y$-dependent
kink solutions are then interpreted as a renormalization group (RG)
flow. The small $y$ (large $a(y)$, hence large UV cutoff
$\Lambda_{UV}$) regime is the ``UV region'', the large $y$ (small
$a(y$), small $\Lambda_{UV}$) regime the ``IR region'' (cf. Fig. 1).
The 4-d theory at small $y$ is always ${\cal N} = 4$ $SU(N)$ $YM$; the
4-d theory at large $y$ can be identified by its unbroken symmetries:
e.g. ``Higgsed'' ${\cal N} = 4$ $YM$, or ${\cal N} = 1$ $YM$, or even
${\cal N} = 0$ $YM$. In any case the 4-d theory at large $y$ is a CFT
(with vanishing $\beta$-function) as long as the 5-d space is AdS$_5$
for large $y$; the examples with ${\cal N} = 0$ supersymmetry
correspond, however, to non-unitary CFTs (the kink-solution does not
end up in a local minimum of the scalar potential for $y \to +
\infty$). \par

On any ``background'' ($y$-dependent, but $x^i$-independent scalars and
metric) one can study ``fluctuations'': $x^i$-dependent (and
$y$-dependent) solutions of the equations of motion of scalars, vectors
$A_i$, and the $(i,j)$ components of the 5-d graviton, with prescribed
boundary-values at some small $y = y_1$. Since the 5-d space is (again)
AdS$_5$ for large $y$, one imposes again decreasing wave functions for
large $y$, and the boundary values at $y_1$ determine again uniquely
the solutions for all $y$. As before, one starts with the linearized
equations of motion (in the fluctuations), and studies, possibly,
higher orders in the fluctuations iteratively. \par

 Then, as before, one computes $S_1(\phi_1)$, the bulk action
integrated from $y = y_1$ to $y = + \infty$. This functional is at
least quadratic in $\phi_1$, and allows to obtain the 2 (or
higher)-point functions of the associated operators. \par

Third, with some courage, one can look for non-trivial background
solutions of $d = 5$ ${\cal N} = 8$ supergravity, which do not behave
like AdS$_5$ for large $y$: If the warp-factor $a(y)$ vanishes for
some finite $y$, the corresponding 4-d theory, in the infra-red, is
some non-conformal theory. \par

Support for this conjecture arises, e.g., from the corresponding
2-point function of the dilaton $\varphi$. The associated 4-d operator
is the trace of the energy-momentum tensor, which contains the
operator $F_{\mu \nu} F^{\mu \nu}$. If one isolates the leading
non-analytic behavior of $S_1(\varphi_1)$ to ${{\cal O}}(\varphi_1^2)$
as in eqs. (\ref{2.20e}), (\ref{2.21e}) above, one obtains no
logarithmic cut in $p^2$, but a sequence of (Regge-) poles in $p^2$.
These are interpreted as a glueball Regge trajectory, since glueballs
would couple to the corresponding operator. Thus one hopes to describe
a confining 4-d gauge theory. \par

 However, the vanishing warp-factor $a(y)$ at finite $y$ implies, in
general, a naked singularity in the 5-d bulk. Most importantly, the
divergent 5-d curvature at such a singularity leads to a breakdown of
the ``supergravity approximation'', i.e. the possibility to neglect
(stringy) higher powers of $\alpha '$ or higher powers of the
curvature tensor in the bulk action. \par

One possibility is to replace the bulk action near the singularity by
the action of the full string theory \cite{J1}. In \cite{G2} it is
proposed  that naked singularities are allowed (and the
cor\-res\-pon\-ding boundary conditions on the fields can be derived),
if they can be approached smoothly in the $T \to 0$ limit of a black
hole solution with temperature $T$, which ``hides'' the singularity
behind a horizon (for $T \not = 0$). \par

It should have become clear that the search for $y$-dependent solutions
of the coupled scalar/gravitational equations of motion plays an
important role in this framework, as well as the computation of the
$y$-integrated bulk action. Also the interpretation of the
$y$-dependence as an RG flow merits further clarification: The
equations of motion contain obviously second derivatives in $y$,
whereas RG equations are generally first order equations. To both ends
a first order formulation of the classical bulk dynamics, i.e. a
Hamilton-like approach, proves to be helpful. This will be the subject
of the next section. 

\mysection{Hamiltonian constraint and RG equations} 
\hspace*{\parindent} The approach developed in the present section is
based, to a large extent, on \cite{B1}. Subsequently we will denote all
degrees of freedom in the bulk, scalars or components of the graviton,
by $q_{\alpha}(y,x^i)$. Furthermore we denote $y$-derivatives by
primes, $\partial_y q_{\alpha} = q'_{\alpha}$. Concerning the 5-d
Lagrangian in the bulk we assume, that all second derivatives stemming
from the curvature scalar have been removed by partial integrations,
and that the corresponding total derivative terms are omitted resp.
cancelled by previously added boundary terms \cite{G1}. Thus we can
write  

\beq \label{3.1e} {\cal L}_5 = {\cal L}_5 \left ( q_{\alpha},
\partial_i q_{\alpha}, q'_{\alpha} \right ) \quad . \eeq

As in section 2 we are interested in the action $S_{1,2}$ involving a
$y$-integration from $y = y_1$ to $y = y_2$:

\beq \label{3.2e} S_{1,2} = - \int_{y_1}^{y_2} dy \int d^4x \ {\cal
L}_5\left ( q_{\alpha}, \partial_i q_{\alpha}, q'_{\alpha} \right )
\eeq

\noi where the fields $q_{\alpha}$ are solutions of the equations of
motion with prescribed boun\-da\-ry values at $y = y_1$ and $y = y_2$.
Now we apply the action principle of classical mechanics, i.e. we
consider the variation of $S_{1,2}$ with fluctuations $\delta
q_{\alpha}$, $\delta \partial_i q_{\alpha}$, $\delta q'_{\alpha}$
around the solutions of the equations of motion:

\beq \label{3.3} \delta S_{1,2} = - \int_{y_1}^{y_2} dy \int d^4x
\left ( {\delta {\cal L}_5 \over \delta q_{\alpha}} \ \delta
q_{\alpha} + {\delta {\cal L}_5 \over \delta \partial_i q_{\alpha}} \
\delta \partial_i q_{\alpha} + {\delta {\cal L}_5 \over \delta
q'_{\alpha}} \ \delta q'_{\alpha} \right ) \quad .\eeq

As usual one writes the variations of derivatives of $q_{\alpha}$ as
derivatives of $\delta q_{\alpha}$ and uses partial integration. Then
one uses that $q_{\alpha}$ solves the equations of motion,

\beq \label{3.4e} {\delta {\cal L}_5 \over \delta q_{\alpha}} -
\partial_i \ {\delta {\cal L}_5 \over \delta \partial_i q_{\alpha}} -
\partial_y \ {\delta {\cal L}_5 \over \delta q'_{\alpha}} = 0 \quad , 
\eeq

\noi and that the space is assumed to be compact in the 
$x^i$-directions. Hence boun\-da\-ry terms arise only from the partial
integration of $\partial_y \delta q_{\alpha}$, and one obtains

\beq \label{3.5e} \delta S_{1,2} = \left . \int d^4x {\delta {\cal
L}_5 \over \delta q'_{\alpha}} \right |_{y_1} \delta q_{\alpha} (y_1,
x^i) - \left . \int d^4x {\delta {\cal L}_5 \over \delta q'_{\alpha}}
\right |_{y_2} \delta q_{\alpha}(y_2, x^i) \quad . \eeq    

\noi Let us now concentrate on the dependence of $S_{1,2}$ on the
boundary values of $q_{\alpha}$ at $y_1$, which we denote by
$\widehat{q}_{\alpha}$: 

\beq \label{3.6e} q_{\alpha}(y_1, x^i) \equiv
\widehat{q}_{\alpha}(x^i) \quad . \eeq

\noi From (\ref{3.5e}) one obtains

\beq \label{3.7e} \left . {\delta S_{1,2} \over \delta
\widehat{q}_{\alpha}} = {\delta {\cal L}_5 \over \delta q'_{\alpha}}
\right |_{y_1} \quad . \eeq

\noi Next we use that the 5-d (classical) Lagrangian ${\cal L}_5$ can
generally be written as

\bminiG{3.8e} \label{3.8ae} {\cal L}_5 \left (q_{\alpha}, \partial_i
q_{\alpha}, q'_{\alpha} \right ) = {1 \over 2} q'_{\alpha} \ {\cal
G}^{\alpha \beta}(q) \ q'_{\beta} + \widetilde{\cal L}_5 \quad ,
\eeeq   \beeq \label{3.8be} \widetilde{\cal L}_5 = {1 \over 2} \
\partial_i q_{\alpha} \ \widetilde{\cal G}^{\alpha \beta}(q) \
\partial_i q_{\beta} + V(q) \quad . \emini

\noi Thus one finds

\beq \label{3.9e} {\delta {\cal L}_5 \over \delta q'_{\alpha}} = {\cal
G}^{\alpha \beta}(q) \ q'_{\beta} \eeq

\noi or, from (\ref{3.7e}) (using (\ref{3.9e}) at $y = y_1$, where
(\ref{3.6e}) holds),

\beq \label{3.10e} \widehat{q}{\, '}_{\alpha} = {\cal G}^{-1}_{\alpha
\beta}(\widehat{q}) {\delta S_{1,2} \over \delta \widehat{q}_{\beta}}
\eeq

\noi This relation will be used below. 

Now we turn to the Hamiltonian constraint. First we note that a solution
$q_{\alpha}$ of the equations of motion, with boundary values at $y_1$
and $y_2$, will be of the general form $q_{\alpha}(y,q(y_1), q(y_2))$.
Inserting the solutions into ${\cal L}_5$ in (\ref{3.2e}), $S_{1,2}$
appears to be of the form

\beq \label{3.11e} S_{1,2} = S_{1,2} \left ( y_1, y_2, q(y_1), q(y_2)
\right ) \quad . \eeq 

However, since ${\cal L}_5$ contains gravity, invariance under general
coordinate transformations guarantees that $S_{1,2}$ does not depend
explicitely on $y_1$, $y_2$; reparametrizations of $y$ can be
compensated for by corresponding variations of the metric (and the
fields). From (\ref{3.11e}) this implies

\beq \label{3.12e} 0 = {\partial S_{1,2} \over \partial y_1} = {d
S_{1,2} \over dy_1} - \int d^4x \ {\delta S_{1,2} \over \delta
q_{\alpha}(y_1, x^i)} \ q'_{\alpha}(y_1, x^i) \quad .\eeq

\noi On the other hand, using (\ref{3.2e}) the total derivative of
$S_{1,2}$ w.r.t. $y_1$ is given by ${\cal L}_5$ at $y = y_1$. Using 
(\ref{3.2e}) and (\ref{3.6e}) in the first line, and (\ref{3.7e}) in the
second line, one obtains  

\bea \label{3.13e}  0 &=& \int d^4 x \left \{ {\cal L}_5 (\widehat{q})
- {\delta S_{1,2} \over \delta \widehat{q}_{\alpha}} \
\widehat{q}_{\alpha}{\, '} \right \} \nn \\ &=& \int d^4 x \left \{
{\cal L}_5 (\widehat{q}) - {\delta {\cal L}_5 (\widehat{q}) \over
\delta \widehat{q}_{\alpha}{\, '}} \ \widehat{q}_{\alpha}{\, '} \right
\} \equiv \int d^4 x \ {\cal H}_5 (\widehat{q}) \quad . \eea

\noi This is a version of the Hamiltonian constraint (${\cal H} = 0$ on
solutions) in general relativity. Here, however, ${\cal H}_5$ is the
generator of translations in $y$ and not, as usual, in $t$. Actually,
also a local version (in $x^i$) of (\ref{3.13e}) can be derived
\cite{B1}, if one would allow for a $x^i$-dependence of the boundary
value $y_1$ (which would require, however, a $x^i$-dependent metric
$\eta_{ij}$ in (\ref{2.2e})). \par

If one uses (\ref{3.8ae}) and (\ref{3.10e}) in the first identity in
(\ref{3.13e}), it can be rewritten as

\beq \label{3.14e} 0 = \int d^4 x \left \{ - {1 \over 2} \ {\delta
S_{1,2} \over \delta \widehat{q}_{\alpha}} \ {\cal G}^{-1}_{\alpha
\beta}(\widehat{q})\  {\delta S_{1,2} \over \delta
\widehat{q}_{\beta}} + \widetilde{\cal L}_5(\widehat{q}) \right \}
\quad . \eeq

This version of the Hamiltonian constraint allows to obtain
informations on the functional form of $S_{1,2}$ in terms of the
component $\widetilde{\cal L}_5$ of the bulk Lagrangian, cf.
(\ref{3.8be}). However, (\ref{3.14e}) does not lead to any constraints
beyond the equations of motion; it is just a convenient way of
expressing their consequences. \par

Let us return to the parametrization (\ref{2.2ae}) of the metric in the
bulk. In addition we confine ourselves to $x^i$-independent
configurations of scalar fields $\varphi_i$. On these configurations
the formal results (\ref{3.10e}) and (\ref{3.14e}) can be used most
easily in order to construct solutions of the equations of motion. Now 
the bulk Lagrangian reads

\beq \label{3.15e} {\cal L}_5 = a^4b \left \{ - {6 \over
a^2b^2\kappa_5^2}\ a'^{\, 2} + {1 \over 2b^2} \ \varphi '_i \ {
G}^{ij}(\varphi ) \ \varphi '_j + V(\varphi ) \right \} \eeq

\noi where ${G}^{ij}(\varphi )$ denotes a sigma model metric, and
$V(\varphi )$ includes a possible cosmological constant. With
$q_{\alpha} = \{a, \varphi_i\}$ ($b'$ does not appear in ${\cal L}_5$
or the Hamiltonian ${\cal H}_5$) one can read off ${\cal G}^{\alpha
\beta}$, as defined in (\ref{3.8ae}), from (\ref{3.15e}):  ${\cal
G}^{aa} = - 12 a^2/b \kappa_5^2$ (where the indices $a$ of ${\cal
G}^{aa}$ correspond to the warp factor $a(y)$) and ${\cal G}^{ij} =
(a^4/b)G^{ij}$. Thus eqs. (\ref{3.10e}) become 

\beq \label{3.16e} \widehat{a}' = -{\widehat{b} \kappa_5^2 \over 12
\widehat{a}^2} \ {\partial S_{1,2} \over \partial \widehat{a}} \quad ,
\qquad \widehat{\varphi} '_i = {\widehat{b} \over \widehat{a}^4} \ {
G}_{ij}^{-1} (\widehat{\varphi}) {\partial S_{1,2} \over \partial
\widehat{\varphi}_j} \quad , \eeq 

\noi where the hats indicate again the fields at $y = y_1$. With ${\cal
G}^{\alpha \beta}$ as above, and $\widetilde{\cal L}_5(\widehat{q}) =
V(\widehat{q}) = \widehat{a}^4\widehat{b}V(\widehat{\varphi})$, 
equation
(\ref{3.14e}) assumes the form, omitting the $d^4x$-integral and
dividing by $\widehat{b}$, 

\beq \label{3.17e} 0 = {\kappa_5^2 \over 24\widehat{a}^2} \left (
{\partial S_{1,2} \over \partial \widehat{a}} \right )^{2} - {1 \over
2\widehat{a}^4} \ {\partial S_{1,2} \over \partial
\widehat{\varphi}_i} \ {G}^{-1}_{ij} (\widehat{\varphi}) {\partial
S_{1,2} \over \partial \widehat{\varphi}_j} + \widehat{a}^4 \
V(\widehat{\varphi}) \quad .\eeq

Reparametrization invariance in $d = 4$ suggests the following ansatz
for $S_{1,2}$ proportional to $\sqrt{-g_4(y_1)} = \widehat{a}^4$: 

\beq \label{3.18e} S_{1,2} = \widehat{a}^4 W(\widehat{\varphi}) +
\dots  \eeq

\noi where the dots denote terms independent of $\widehat{a}$,
$\widehat{\varphi}$ arising, possibly, from the upper end $y = y_2$ of
the $y$-integration. With this ansatz eqs. (\ref{3.16e})
become 

\beq \label{3.20e} {\widehat{a}' \over \widehat{a}} = -{ \widehat{b}
\kappa_5^2 \over 3} \ W \quad , \qquad \widehat{\varphi}'_i =
\widehat{b} \ {G}_{ij}^{-1} \ W_{,j} \quad , \eeq

\noi and eq. (\ref{3.17e}) can be brought into the form

\beq \label{3.19e} {2\kappa_5^2 \over 3} \ W^2 - {1 \over 2} \ W_{,i} \
{G}_{ij}^{-1} \ W_{,j} + V(\widehat{\varphi}) = 0 \eeq

\noi with $W_{,i} = \partial W/\partial \widehat{\varphi}_i$. 

In \cite{F1} (in the metric (\ref{2.3ae}), where $\widehat{b} =
1$, and with ${G}_{ij} = \delta_{ij}$ and different conventions in the
gravitational sector) eqs. (\ref{3.20e}) and (\ref{3.19e}) have been
proposed as a ``short cut'' towards the search for solutions of the
equations of motion in the bulk: Instead of solving the coupled second
order differential equations (\ref{3.4e}) for $a(y)$, $\varphi_i(y)$
one first tries to find a ``superpotential'' $W(\varphi )$ which solves
(\ref{3.19e}) (with $V(\varphi )$ given). Then one is left with the
integration of the remaining first order equations (\ref{3.20e}). The
number of integration constants matches: For each scalar field there is
one from (\ref{3.19e}) and one from the second of eqs. (\ref{3.20e}),
in agreement the analysis in section 2. The combined Einstein equations
and Bianchi identities also allow for just one integration constant for
the warp factor, in agreement with the first of eqs. (3.20). \par

It should be emphasized, however, that not all solutions can be written
in the form (\ref{3.18e})-(\ref{3.19e}). A counter example is given by
the iterative solution (\ref{2.9e}) in section 2, with $C_1(p) =
c_1p^{\nu}$, $C_2(p) = c_2p^{-\nu}$ and $p \to 0$, if both $c_1$ and
$c_2$ are non-zero. Only for $c_1 = 0$ or $c_2 = 0$ the solution can be
written in the form (\ref{3.19e}). Generally, solutions of the form
(\ref{3.18e})-(\ref{3.19e}) have the particular property of preserving
${\cal N} = 1$ supersymmetry \cite{F2,C1,F1}. \par

Let us now return to 5-d brane universes, and re-install branes at
$y_1$, $y_2$ with actions $S^{(1)}(a,b,\varphi )$ and
$S^{(2)}(a,b,\varphi )$ respectively. These imply jump conditions
\cite{BD1,RS1,RS2} to the right of $y_1$ of the form

\beq \label{3.21e} {a' \over a}(y_1) = {1 \over 2} \left [ - {b \over
12a^3}\ {\partial S^{(1)} \over \partial a} \right ]_{y_1} \quad ,
\quad \varphi '_i(y_1) = {1 \over 2} \left [ {b \over a^4} \ {
G}_{ij}^{-1} {\partial S^{(1)} \over \partial \varphi_j} \right ]_{y_1}
\eeq

\noi and jump conditions to the left of $y_2$ of the form

\beq \label{3.22e} {a' \over a}(y_2) = {1 \over 2} \left [  {b \over
12a^3}\ {\partial S^{(2)} \over \partial a} \right ]_{y_2} \quad ,
\quad \varphi '_i(y_2) = {1 \over 2} \left [ - {b \over a^4} \ {
G}_{ij}^{-1} {\partial S^{(2)} \over \partial \varphi_j} \right
]_{y_2} \quad . \eeq 

\noi Here we assumed orbifold boundary conditions. If, instead, the
branes indicate ``ends of the world'', the factors 1/2 in
(\ref{3.21e}) and (\ref{3.22e}) have to be omitted. \par

Clearly, eqs. (\ref{3.21e}) and (\ref{3.22e}) have to be consistent
with eq. (\ref{3.20e}) at $y = y_1$, $y_2$, respectively, in order to
allow for a global solution. If these equations are inconsistent,
$x^i$-independent solutions for $a$, $b$ and $\varphi_i$ do not exist;
in particular this implies an $x^i$-dependent warp factor $a(y, x^i)$
on ``our'' brane in contradiction to the (practically) static and
homogeneous observed universe. The argument can be turned around:
assuming a static and homogeneous universe, and 

\beq \label{3.23e} S^{(1)} = a^4 W^{(1)} \quad , \qquad S^{(2)} =
a^4 W^{(2)} \quad , \eeq

\noi one can derive $W^{(2)} = - W^{(1)} = \pm 2W$ where $W$ has to
satisfy (\ref{3.19e}) \cite{E1}. \par

Supersymmetry is not involved in deriving these constraints, which
coincide with the ones employed in \cite{RS1,RS2}, but supersymmetry
helps to satisfy them (see, e.g., \cite{C1,L1}). \par

Finally we turn to the interpretation of $S_{1,2}(\widehat{\varphi})$
as the generating functional of connected Green functions of composite
operators, and of eqs. (\ref{3.16e}) as RG equations for the
``sources'' (= couplings) $\widehat{a}$, $\widehat{\varphi}_i$ at $y =
y_1$. As already stated in section 2 it is not reasonable to identify
$y$ directly with a RG scale; as in eq. (\ref{2.18e}) we should rather
associate an UV cutoff (which we identify with a RG scale subsequently;
the following RG equations have to be interpreted correspondingly) with
the warp factor $a(y)$. With

\beq \label{3.24e} \Lambda_{UV} {\partial \over \partial\Lambda_{UV}} =
a(y) {\partial \over \partial a(y)} \eeq

\noi and

\beq \label{3.25e} {\partial \over \partial y} = a' {\partial \over
\partial a} = {a' \over a} \Lambda_{UV} {\partial \over
\partial\Lambda_{UV}} \quad , \eeq

\noi we can rewrite the second of eqs. (\ref{3.16e}) as (omitting the
hats in the following)

\bea \label{3.26e} \Lambda_{UV} {\partial \varphi_i \over \partial
\Lambda_{UV}} &=& - {12 \over \kappa_5^2} \ {G}_{ij}^{-1} \ {\partial
S_{1,2} \over \partial \varphi_j} \left ( a {\partial S_{1,2} \over
\partial a} \right )^{-1} \nn \\ &\equiv & \beta_i(\varphi) \quad .
\eea

\noi In particular, with the ansatz (\ref{3.18e}) for $S_{1,2}$, one
obtains

\beq \label{3.27e} \beta_i(\varphi ) = - {3 \over \kappa_5^2} \ {
G}_{ij}^{-1}( \varphi ) {W_{,j} \over W} \quad . \eeq

\noi Hence the holographic RG-flow is derived from a potential
\cite{B1}, i.e. it is proportional to the gradient of a c-function $W$
\cite{F1}. \par

However, $S_{1,2}$ does not necessarily have to be of the form
(\ref{3.18e}). Let us assume, following \cite{B1,V1}, that $S_{1,2}$
is given by

\beq \label{3.28e} S_{1,2} = a^4 \left ( W(\varphi ) +
\widetilde{W}(a,y) \right ) \eeq

\noi where $\widetilde{W}$ is a small correction. Inserting
(\ref{3.28e}) into (\ref{3.17e}) and using (\ref{3.19e}) one obtains,
to first order in $\widetilde{W}$,

\beq \label{3.29e} \left ( a {\partial \over \partial a} + 4 - {3
\over \kappa_5^2} \ {W_{,i} \over W} \ {G}_{ij}^{-1} \ {\partial \over
\partial \varphi_j} \right ) \widetilde{W} = 0 \quad . \eeq

\noi With (\ref{3.24e}) and (\ref{3.27e}) this becomes indeed an RG
equation for $\widetilde{W}$:

\beq \label{3.30e} \left ( \Lambda_{UV} {\partial \over \partial
\Lambda_{UV}} + 4 + \beta_i {\partial \over \partial \varphi_i} \right
) \widetilde{W} = 0 \eeq

Further consequences of the RG interpretation of the Hamiltonian
approach to the $y$-dependence are derived in \cite{B1}. (It should be
noted, however, that the step from (\ref{3.30e}) towards a RG equation
involving an infra-red scale $\mu$ with $\Lambda_{UV} \partial /
\partial \Lambda_{UV} = - \mu \partial / \partial \mu$ holds only for
scale invariant theories; otherwise $\mu$ has no physical
significance). 

In \cite{V2} an equation is derived, which ressembles Polchinski's
exact renormalization group equation \cite{P1}. To this end one
defines

\beq \label{3.31e} S = S_{UV} + S_{IR} = S_{y_1,y} + S_{y,y_2} \eeq

\noi with, possibly, $y_1 \to -\infty$ (in the metric (\ref{2.2ae}) with
$b(y) = 1$). Here

\beq \label{3.32e}
S_{UV} = S_{y_1,y} \eeq

\noi is interpreted as an effective action in the Wilsonian sense, where
degrees of freedom with momenta $p^2$ with $M^2a^2(y_1) > p^2 > M^2
a^2(y)$ have been integrated out. ($M^2$ is some fundamental scale.)
Similarly, 

\beq \label{3.33e}
S_{IR} = S_{y,y_2}, \eeq

\noi with $a(y_2)=0$, corresponds to an effective action involving the
path integral over modes with $M^2 a^2(y) > p^2 > 0$. Hence $S$ is the
full quantum effective action, which is splitted into its UV and IR
part in (\ref{3.31e}). The "split-point" corresponds to a scale $\mu$
with $\mu^2 = M^2 a^2(y)$. \par

In analogy to (\ref{3.26e}) one defines $\beta$-finctions as

\beq \label{3.34e} \beta_i = -{12 \over \kappa_5^2} {G}^{-1}_{ij}
{\partial S_{UV} \over \partial \varphi_j} \left ( a(y) {\partial
S_{UV} \over \partial a(y)} \right )^{-1}\ .  \eeq

Returning to ${\cal G}^{ij} = (a^4 / b)G^{ij}$  and defining

\beq \label{3.35e}
\gamma = -{b \kappa_5^2\over 12 a^4}
 \left(a(y) {\partial S_{UV} \over \partial a(y)} \right)
\eeq

\noi eq. (\ref{3.34e}) can be expressed as

\beq\label{3.36e}
\gamma \beta_i = {\cal G}^{-1}_{ij}{\partial S_{UV} \over \partial
\varphi_j(y)}\ . \eeq

\noi (The minus sign in (\ref{3.35e}) is related to the fact that now
we take the variation of $S_{UV}$ at the upper limit of the $y$
integral.) Defining the $\beta$ function of the warp factor as
$\beta_{a} = a$ eq. (\ref{3.36e}) holds again for all fields
$q_{\alpha} = \{a, \varphi_i\}$:

\beq\label{3.37e} \gamma \beta_{\alpha} = {\cal G}^{-1}_{\alpha
\beta}{\partial S_{UV} \over \partial q_{\beta}(y)}\ . \eeq

\noi Multiplying (\ref{3.37e}) by $\partial S_{UV} / \partial
q_{\alpha}(y)$ one obtains 

\beq\label{3.38e} \gamma \beta_{\alpha} {\partial S_{UV} \over \partial
q_{\alpha}} = {\partial S_{UV} \over \partial q_{\alpha}}  {\cal
G}^{-1}_{\alpha \beta} {\partial S_{UV} \over \partial q_{\beta}} \ .
\eeq

Eq. (\ref{3.38e}) shows some formal similarity to Polchinski's exact
renormalization group equation \cite{P1}. A similar equation can be
derived for $S_{IR}$. In \cite{V2} (below eq. (24)) the relevance of
(\ref{3.38e}) for $S = S_{UV} + S_{IR}$ is emphasized. Since, however,
classical solutions $q_{\alpha}$ extremize the full effective action
$S$ (the contributions from $S_{UV}$ and $S_{IR}$ in (\ref{3.31e})
cancel) it becomes now a trivial identity.

\mysection{Vanishing cosmological constant}  
\hspace*{\parindent} As discussed before, the problem of the vanishing
4-d cosmological constant in a 5-d brane universe can be phrased as the
problem of obtaining  $x^i$-independent solutions for the warp factor
$a(y)$. Given the jump conditions at the branes (and, in addition,
continuity of the fields across the branes) one typically ends up with
more constraints than available integration constants. Thus, as in
2-brane universes with orbifold boundary conditions \cite{BD1,RS1,RS2},
the parameters of the actions in the bulk and on the branes have to be
fine tuned relatively to each other.
\par

Recent proposals to avoid such fine tunings are: In \cite{A1} a
1-brane/2-bulks scenario (with orbifold-like boundary conditions) is
considered, where a scalar field -~which couples like a dilaton~-
ensures the compatibility of the jump conditions with the bulk
equations of motion. In \cite{K1} a 1-brane/2-bulks scenario (without
orbifold-like boundary conditions) is considered, where the number of
constraints does not exceed the number of integration constants. In
both cases, however, a naked singularity (where $a(y) = 0$) is
encountered at finite values of $y$. \par

The proposal in \cite{V2,V3,V1} is quite different: It is based on a
2-branes/1-bulk scenario which, a priori, seems to require fine tunings
among the actions on the branes and in the bulk. However, the RG
interpretation of the $y$-dependence (or $a(y)$-dependence) of
integrated bulk action is taken literally, motivated by the
AdS/CFT correspondence: First, the small-$y$ region (where $a(y)$ is
large) is identified with the UV regime of field theory/string theory.
Since supersymmetry is valid in this regime, an action on a brane at
small $y_1$ allows for jump conditions consistent with the bulk action.
(In the approximation of $x^i$-independent field configurations as
considered near the end of section 3, the superpotential $W^{(1)}$ in
(\ref{3.23e}) is related to the bulk potential $V$ via (\ref{3.19e}),
due to supersymmetry at the ``large scale'' $y_1$. Hence eqs.
(\ref{3.20e}) and eqs. (\ref{3.21e}) (without factors 1/2) are
consistent due to supersymmetry.) Then one considers $S_{y_1,y}$ given
by

\beq \label{4.1e} S_{y_1,y} = - \int_{y_1}^y dy \int d^4x \ {\cal
L}_5 \quad . \eeq      

\noi which is interpreted (as discussed below eq. (\ref{3.31e})) as a
Wilsonian effective action. Recall that, for $y \to y_2$ with $a(y_2) =
0$, $S_{y_1,y_2}$ cor\-res\-ponds to the full quantum effective action.
It is assumed that at some intermediate scale dynamical supersymmetry
breaking takes place. The brane where we live on is situated at $y_2$.
In agreement with previous AdS/FT results the integrated action
$S_{y_1,y_2}$ corresponds to a non-conformal field theory (as our
world), since the bulk space-time near the naked singularity at $y_2$
is not AdS$_5$. The fact that the jump conditions match at $y_2$ is
considered as a tautology: The action $S^{(2)}$ on the brane 2 is the
full quantum effective action (generating a possible cosmological
constant on brane 2), but $S_{y_1,y_2}$ is the {\bf same} effective
action: 

\beq \label{4.2e} S_{y_1,y_2} = S^{(2)} \quad .\eeq 

Hence a combined solution of (\ref{3.16e}) and
(\ref{3.22e}) (after a trivial change of sign of $S^{(2)}$, and without
the factors $1/2$) with $a(y)$ independent from $x^i$, for all $y$
including the IR regime $y_2$, is possible. Although $a(y)$ is no
longer a component of the metric from the 4-d point of view, but rather
a source for an operator, this argument indicates that a flat brane --
consistent with our (practically) static and homogenous universe -- is
a solution of the combined eqs. of motion and jump-conditions at any
value of $y$. \par

The equality (\ref{4.2e}) is supported by the results in
\cite{B1,V2,V3,V1} and sketched near the end of the previous section
which indicate, that $S_{y_1,y}$ satisfies a RG-like flow equation with
respect to $a(y)$ which is of the same form as the RG equation
satisfied by a cutoff quantum effective action $S^{(2)}$. In
\cite{V2,V3,V1,KV} arguments are put forward which should support the
identification of $S_{y_1,y}$ with the quantum effective action beyond
the previously employed approximations, as $x^i$-independent field
configurations and, notably, a classical action in the bulk. To our
opinion it still remains to be shown, however, whether a cutoff quantum
effective action -~with some concrete definition of the cutoff, which
also remains to be found~- can be written as an integrated bulk action,
without contradicting any of our knowledge of local quantum field
theory. After all these concepts should remain valid far below the
string scale.     

\vskip 2 cm
\noi {\Large\bf Acknowledgements}

It is a pleasure to thank the members of the LPT Orsay for discussions
and helpful comments.

\end{document}